# Influence of static electric fields on an optical ion trap

Christian Schneider,[1,2] Martin Enderlein,[1,2] Thomas Huber,[1,2] Stephan Dürr,[1] and Tobias Schaetz[1,2]

[1]*Max-Planck-Institut für Quantenoptik, Hans-Kopfermann-Str. 1, 85748 Garching, Germany*
[2]*Albert-Ludwigs-Universität Freiburg, Physikalisches Institut, Hermann-Herder-Str. 3, 79104 Freiburg, Germany*

We recently reported on a proof-of-principle experiment demonstrating optical trapping of an ion in a single-beam dipole trap superimposed by a static electric potential [1]. Here, we first discuss the experimental procedures focussing on the influence and consequences of the static electric potential. These potentials can easily prevent successful optical trapping, if their configuration is not chosen carefully. Afterwards, we analyse the dipole trap experiments with different analytic models, in which different approximations are applied. According to these models the experimental results agree with recoil heating as the relevant heating effect. In addition, a Monte-Carlo simulation has been developed to refine the analysis. It reveals a large impact of the static electric potential on the dipole trap experiments in general. While it supports the results of the analytic models for the parameters used in the experiments, the analytic models cease their validity for significantly different parameters. Finally, we propose technical improvements for future realizations of experiments with optically trapped ions.

## INTRODUCTION

Since the first trapping of ions in RF traps by Paul et al. in 1958 [2] and the first trapping of atoms in optical traps by Chu et al. in 1986 [3], these trapping techniques have substantially advanced various fields of research. This includes, for example, quantum information processing [4], quantum metrology [5, 6], cold chemistry [7], and quantum degenerate gases [8]. However, trapping ions optically has been reported only recently [1]. A reason is that the Coulomb interaction leads to a high sensitivity to stray electric fields which might also account for the prejudice that ions would not be trappable in comparatively weak optical potentials. Potential depths of optical dipole traps, for example, are on the order of $10^{-3}$ K, while conventional RF traps have depths on the order of $10^4$ K.

An overview of the setup for optical trapping [1] is shown in Fig. 1. The basic experimental sequence is the following: Initially, a single ion is loaded and laser-cooled in a conventional, linear RF trap. Subsequently, a dipole trap beam is superimposed with the ion and the RF potential is ramped down to zero. Finally, after the dipole trap duration $T_{\text{optical}}$, the RF potential is ramped up again and the ion is detected in case of successful former optical trapping. During all steps a static electric potential is retained, such that the total potential consists of the dipole trap and a static electric potential. Firstly, the static electric potential is required to compensate stray electric fields and, secondly, it prevents the ion from leaving the dipole trap in the propagation direction of the beam $\hat{k}$. Its focussing effect along the $z$-axis comes at the price of a defocussing effect in at least one radial direction, which has to be overcome by the dipole trap potential. Furthermore, the 45° angle between the $z$-axis and $\hat{k}$ results in an asymmetric total potential.

In this paper we first discuss the experimental procedures and peculiarities. While the RF potential is still on, the axial potential shows strong, unexpected anhar-monicities. Using ions as sensors we measure, quantify and quantify the anharmonicities and deduce consequences for trapping ions in the dipole trap. We also describe the procedure for compensating stray electric fields, such that the resulting forces are smaller than the maximal forces in the dipole trap. Likely reasons for these stray electric fields are presented. Afterwards, the ramp-down procedure of the RF potential for the transfer of the ion into the dipole trap is described and discussed with the help of the stability diagram of a RF trap. Again, the influence of the static electric potential is highlighted. In the following, we describe the dipole trap setup. The method to adjust the polarization of the dipole trap beam to a purity of better than 1 : 1000 (intensity) is explained, which is of importance to obtain the maximal potential depth of the dipole trap and minimal heating. Furthermore, a brief discussion of the heating effects in dipole traps is presented revealing that recoil heating is expected to be dominant in our dipole trap beam.

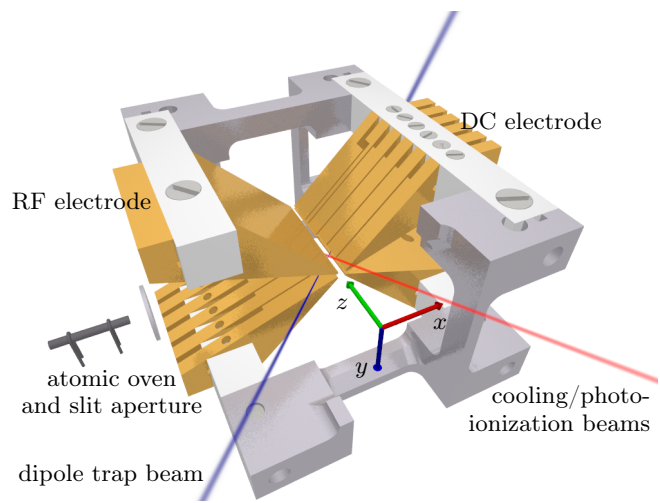

Figure 1. Scheme of the trapping setup (color online). The linear RF trap consists of gold-plated, blade-like electrodes, which are mounted using alumina isolators and stainless steel holders. The DC electrodes are divided into six segments each: three belong to a zone with greater electrode–ion distance and three to a zone with smaller electrode–ion distance. (The latter zone is not used in the current dipole trap experiments.) The atomic oven can evaporate a beam of neutral magnesium atoms mainly in direction of the zone with large electrode–ion distance. One or more atoms are photo-ionized and Doppler cooled in the RF potential using laser beams in the $y$-$z$-plane, which intersect the $z$-axis at an angle of 22.5°. The dipole trap beam propagates in the $x$-$z$-plane and crosses the $z$-axis at an angle of 45°. The fluorescence light from the ion(s) is detected with a CCD camera through a sapphire window above the trap (not shown). Parts of an isolator and a holder are omitted to offer a view on the trapping zone. The arrows of the coordinate system indicate a length scale of 10 mm; the distance between the atomic oven and its aperture is not to scale.

In the following the experimental results are analysed. The analytic model used in Ref. 1 is derived. It assumes several simplifications, but is in agreement with the theoretical expectation of recoil heating as the dominant heating effect. One of the simplifications is the cutoff of the trapping probability distribution in energy space, while in the experiment the cutoff happens in the position or potential energy space, respectively. An improved analytic model is introduced to overcome this simplification. Furthermore, the improved model allows for an estimate of the recoil heating rate as a function of temperature, which was previously assumed to be constant. Still, both analytic models simplify the dipole trap potential as a spherically symmetric harmonic potential. We analyse the asymmetric shape of the potential consisting of dipole trap and static electric potential (including its defocussing effect) and quantify its depth as a function of the parameters of the static electric potential. Our setup offers only limited possibilities for measuring the parameters of the static electric potential, but according to our analysis these partly unknown parameters have a substantial effect on the overall trapping potential. In the best case scenario they effectively reduce the depth of the dipole trap by already 25 %. In the worst case scenario, the overall potential does not have a local minimum, meaning that trapping becomes impossible. To refine the analysis, a Monte-Carlo simulation has been developed, which accounts for the asymmetry of the total potential. Besides several other experimental details, the Monte-Carlo simulation also accounts for the Doppler cooling process in the RF trap and the transfer of the ion into the dipole trap. It is in good agreement with recoil heating as the dominant heating effect and supports the results of the analytic models for the experimental parameters. However, it also reveals a limited validity of the analytic models for significantly different parameters. Finally, we conclude with a discussion of the experimental limitations of our setup, propose improvements, and give an outlook of possible applications for optical trapping of ions.

The article is organized as follows: In Sec. I we discuss the experimental procedures, their peculiarities, and difficulties. In Sec. II, we provide a comparison of the different models used to analyse the experimental results. Finally, in Sec. III, we present concluding remarks.

## I. EXPERIMENTAL PROCEDURE

A single $^{24}$Mg$^+$ ion is created by photo-ionization from a thermal atomic beam and Doppler cooled to a few mK (Doppler cooling limit: 1 mK) in the segmented linear RF trap [9]. We currently use the zone consisting of the three DC segments with the greater electrode–ion distance of $R_0 = 0.8$ mm (Fig. 1). The radial frequencies in the RF trap result from the RF potential and are set to $\omega_{x,y} \approx 2\pi \times 900$ kHz corresponding to a potential depth of $U_{x,y} \approx k_\mathrm{B} \times 10^4$ K, where $k_\mathrm{B}$ denotes the Boltzmann constant.

### A. Axial Confinement by DC Electric Fields

The axial frequency is tuned to $\omega_z \gtrsim 2\pi \times 100$ kHz during the loading process of the RF trap by applying typically positive DC voltages $V_\mathrm{DC,out}$ to the outer segments. The ion has a distance of approximately $Z_0 = 1.5$ mm to the edges of the outer DC segments. In principle, we would expect a focussing effect along the $z$-axis over the complete range of $\pm Z_0$ due to the DC fields and a well depth on the order of $U_z \sim k_\mathrm{B} \times 10^3$ K.

However, the axial potential of the RF trap shows strong anharmonicities, in particular, at a low axial frequency $\omega_z \approx 2\pi \times 45$ kHz, which is chosen during the experimental procedures (i.e., compensation of stray electric fields and trapping in the dipole trap) described later. The images in Fig. 2 show ion chains at different axial frequencies. The symmetry for the high axial frequency indicates that a harmonic confinement is a good approximation for the respective parameters (Fig. 2a), while at the low axial frequency the shift of the center of mass and the different mutual distances between neighbouring ions visualize the axial anharmonicities (Fig. 2b). Due to the anharmonicities it is not possible to trap more than three ions at the low axial frequency chosen for Fig. 2b.

The potential is characterized quantitatively in the following: We extract the (absolute) positions of one, two, and three ions from Fig. 2b and measure the oscillation frequency of the single ion by resonantly exciting its axial motion with a modulation of the voltage of a DC segment. (We do not consider the oscillation frequencies of multiple ions, because the anharmonic potential inhibits sharp resonances then.) The six ion positions and the oscillation frequency are used to fit the parameters $\omega_\mathrm{cub}$,

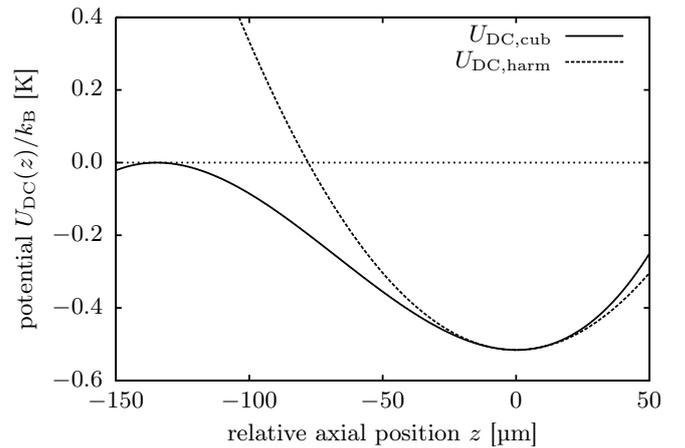

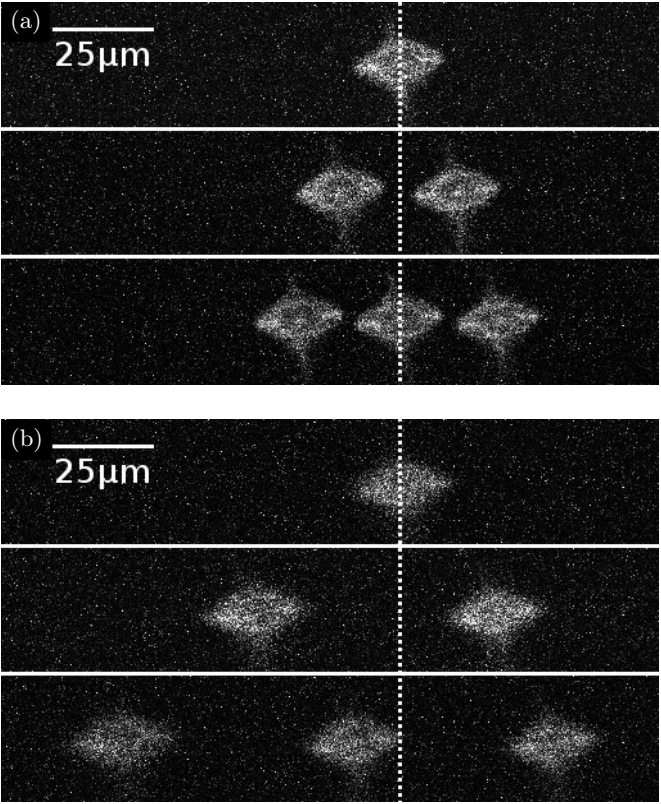

Figure 2. Fluorescence images of one, two, and three ions in the RF trap. The radial frequencies are set to $\omega_{x,y} \approx 2\pi \times 870\,\text{kHz}$, while the axial trap frequency of a single ion amounts to (a) $\omega_z = 2\pi \times (114.0 \pm 0.5)\,\text{kHz}$ and (b) $\omega_z = 2\pi \times (40 \pm 2)\,\text{kHz}$, respectively. The dashed lines indicate the position of a single ion. The objective lens used to image the ions is intentionally defocussed to gain additional information about the ions' positions perpendicular to the imaging plane: The ion images will become larger (smaller), if the ion moves towards (away from) the CCD camera. Figure (b) reveals anharmonicities of the axial potential.

$z_1$, and $z_2$ of

$$U_{\text{cub}}(z) = U_{\text{DC,cub}}(z) + F_{\text{Doppler}}\, z$$
$$= \frac{m}{2} \omega_{\text{cub}}^2 (z - z_1)^2 \left[1 + \frac{2}{3 z_2}(z - z_1)\right] + F_{\text{Doppler}}\, z. \quad (1)$$

Here, $U_{\text{DC,cub}}(z)$ denotes the cubic approximation of the the static electric potential, $F_{\text{Doppler}}$ the $z$-component of the (estimated) force corresponding to the light pressure of the Doppler cooling beams, and $m$ the mass of the ion.

The result of the best fit of $U_{\text{DC,cub}}(z)$ is presented in Fig. 3. The best fit values for $\omega_{\text{cub}}$ and $z_1$ reproduce well the oscillation frequency of a single ion and its position, respectively. The best fit value $z_2 = -135\,\mu\text{m}$ determines the distance of the local maximum to the local minimum and can be interpreted as a measure for the strength of the anharmonicities. The potential has a well depth of $m\omega_{\text{cub}}^2 z_2^2 / 6 = k_B \times 0.5\,\text{K}$.

Figure 3. Axial potential of the RF trap in cubic approximation $U_{\text{DC,cub}}$ and harmonic approximation $U_{\text{DC,harm}}$. The cubic approximation is derived from the positions of one, two, and three ions and the frequency of the axial oscillation of a single ion (Fig. 2b). The light pressure of the Doppler cooling beam is substracted. The axial potential has a maximum at a distance of $135\,\mu\text{m}$ from the minimum and a depth of approximately $k_B \times 0.5\,\text{K}$.

The asymmetric potential is probably due to the loading process that coats the trap electrodes with Mg leading to contact potentials ($-1.44\,\text{V}$ for Mg on Au). This assumption is supported by the fact that a negative voltage of $-1.47\,\text{V}$ has to be applied to one pair of outer DC segments for axial frequencies around $\omega_z \sim 2\pi \times 45\,\text{kHz}$, while the voltage applied to the center segments is close to zero. This can only be explained by confining potentials on length scales smaller than $Z_0$. (The axial frequency due to the RF potential is estimated to $\sim 2\pi \times \text{kHz}$ and can be precluded as reason.) The apparent solution of shuttling the ions into a zone of the RF trap, where the electrodes are not coated by the atomic oven, is not an option in our setup, because the electrode–ion distance is smaller ($R_0 = 0.4\,\text{mm}$) and the dipole trap beam would produce significant stray light.

The strong dependence of the axial potential depth on the axial frequency prevents a further reduction of the static electric potential below $\omega_z \sim 2\pi \times 40\,\text{kHz}$ during the dipole trap experiments. In addition, the ions are prone to loss by collisions with components of the residual gas already at the axial well depth of $k_B \times 0.5\,\text{K}$ leading to a lifetime on the order of minutes. To increase their lifetime the axial frequency is increased to $\omega_z \gtrsim 2\pi \times 100\,\text{kHz}$ after the experimental procedures and during the loading process.

Note that the axial potential can still be well approximated by a harmonic potential while trapping an ion in the dipole trap, because the displacement of the ion from the center of the RF trap remains small. In these experiments an anharmonicity of the total potential is due to the angle of $45\,°$ between the dipole trap beam and the $z$-principal axis of the static electric potential

(cmp. Sec. II B).

## B. Compensation of Stray Electric Fields

In the presence of static electric fields the ion is displaced from the node of the RF potential. If the RF voltage and thus the stiffness of the RF potential is decreased, the ion will be further displaced. Loading the ion into the dipole trap might fail, because the static electric fields can easily exceed the maximal forces of the dipole trap.

Static electric fields have two major contributions: Firstly, the DC voltages applied to the DC segments lead to well controllable fields. Secondly, charges on dielectrics in the surrounding of the RF trap result in stray electric fields, which vary during the experiments. The optimization of the DC voltages and compensation of the stray electric fields is achieved with the ion as a sensor: The RF voltage is gradually ramped down and the ion's displacement is counteracted by appropriate DC voltages applied to the DC segments (see Fig. 1) and an additional wire placed below the RF trap. For each iteration step of the compensation the RF voltage is ramped down only for a duration on the order of 0.5 s to maintain the temperature of the RF circuit. Currently, we are able to perform the compensation without losses down to approximately 15 % of the initial amplitude of the RF voltage and thus to 15 % of the initial frequency ($\omega_{x,y} \approx 0.15 \times 2\pi \times 900\,\text{kHz} = 2\pi \times 135\,\text{kHz}$). However, trapping in the RF trap was not yet possible at $< 10\,\%$ of the initial amplitude. The voltage applied to one of the inner DC segments is fine-tuned on the order of a few 100 µV. Due to the electrode–ion distance of $R_0 = 0.8\,\text{mm}$ and geometric considerations this corresponds to a residual force on the order of $10^{-20}\,\text{N}$ at the position of the ion.

As stated above the RF potential of a finite linear RF trap has a small contribution to the axial confinement. A displacement of the ion during the ramp down of the RF voltage may also be observed in the axial direction. In this case, the axial position of the ion must be adjusted accordingly. However, this has to be done only once, because the node of the RF potential is fixed.

We speculate that there are two main sources of stray electric fields in our setup, some of which require a compensation after each loading of an ion: Firstly, dielectrics in proximity of the ion are charged due to the photoelectric effect by stray laser light. The timescale for the discharge amounts to several minutes. Furthermore, as mentioned in Sec. I A, each loading process further coats the trap electrodes and thereby changes the contact potentials. The resulting changes of the axial potential (cmp. Fig. 3) are not relevant on timescales of weeks and the durations of the experiments presented in Sec. II, respectively.

For quantum simulation experiments in this RF trap [10, 11] the fluorescence modulation technique from Ref. 12 has been used to optimize the DC fields and minimize micromotion, respectively. However, after a micromotion compensation with this technique a reduction of the RF voltage still displaces the ion. We speculate that thermal effects due to other operational parameters of the RF trap in the quantum simulation experiments lead to this discrepancy. That is why the fluorescence modulation technique might be adequate to minimize the micromotion for stiff and constant confinements, but might be less suitable for a compensation of stray electric fields in the case of negligible RF confinements.

## C. Ramping Down the RF Voltage

Besides of the compensation of stray electric fields at the position of the ion, the shape of the static electric potentials is important for a successful trapping in the dipole trap. The DC electric potential can be approximated to quadratic order at the position $\vec{r} = (0,0,0)$ of the ion:

$$U_{\text{DC}}(\vec{r}) = \frac{1}{2} m \left( \omega_{x'}^2 x'^2 + \omega_{y'}^2 y'^2 + \omega_z^2 z^2 \right). \quad (2)$$

It must satisfy Laplace's equation

$$\Delta U_{\text{DC}}(\vec{r}) = 0 \Rightarrow \omega_{x'}^2 + \omega_{y'}^2 + \omega_z^2 = 0. \quad (3)$$

Hence, for real $\omega_z > 0$, at least one out of $\omega_{x'/y'}$ must be imaginary resulting in a defocusing effect in the corresponding direction. Due to the restrictions in the reduction of $\omega_z$ discussed in Sec. I A, the defocusing is comparatively strong in our current setup. The defocusing effect will not only overcome the weak confinement of the dipole trap, if the static electric potential is not adjusted carefully. It also prevents trapping in the RF trap at small RF voltages, which are eventually applied while transferring the ion from the RF into the dipole trap. In the following we discuss this defocusing effect and the ramp-down procedure of the RF voltage with the help of the stability diagram of the RF trap.

The radial motion of the ion in the RF trap (without considering an axial confinement at the moment) is described by Mathieu's equation [13, 14]. Whether trapping is possible, depends on the parameters $q = \frac{2eV_{\text{RF}}}{\Omega_{\text{T}}^2 m R_0^2}$ and $a = \pm a_{x'/y'} = \frac{4eV_{\text{DC,com}}}{\Omega_{\text{T}}^2 m R_0^2}$. Here, $e$ denotes the elementary charge, $\Omega_{\text{T}} = 2\pi \times 56\,\text{MHz}$ the drive frequency, $m \approx 24\,\text{u}$ the mass of the ion, $V_{\text{RF}}$ the amplitude of the RF voltage, and $V_{\text{DC,com}}$ a common DC voltage applied to all DC segments. The coordinates $x'$ and $y'$ correspond to a coordinate system rotated about the $z$-axis ($\hat{e}_{x'/y'} = (\hat{e}_x \mp \hat{e}_y)/\sqrt{2}$ in this context; cmp. Fig. 1). The lowest order stability region for pairs $(q, a)$ is shown in Fig. 4. Carelessly decreasing the RF voltage eventually leads to the loss of the ion before zero voltage is reached and is similar to the scan of a RF massfilter.

However, it will still be impossible to approach arbitrarily small RF voltages, while remaining in the stable region, if the axial confinement and its defocusing effect, respectively, are taken into account. It can be considered by introducing voltages $V_{\mathrm{DC,out}} \neq V_{\mathrm{DC,com}}$ at the outer DC segments. The $a$-parameters of Mathieu's equation change according to $a_{x'/y'} \rightarrow a_{x'/y'} - \kappa_{x'/y'} \frac{4e(V_{\mathrm{DC,out}} - V_{\mathrm{DC,com}})}{\Omega_{\mathrm{T}}^2 m Z_0^2}$, where $\kappa_{x'/y'}$ denote geometry factors [14]. Alternatively, the stable regions for the motion in $x'$- and $y'$-direction, respectively, can be shifted for a specific $V_{\mathrm{DC,out}}$, while the $a$-parameters remain unchanged (see Fig. 5).

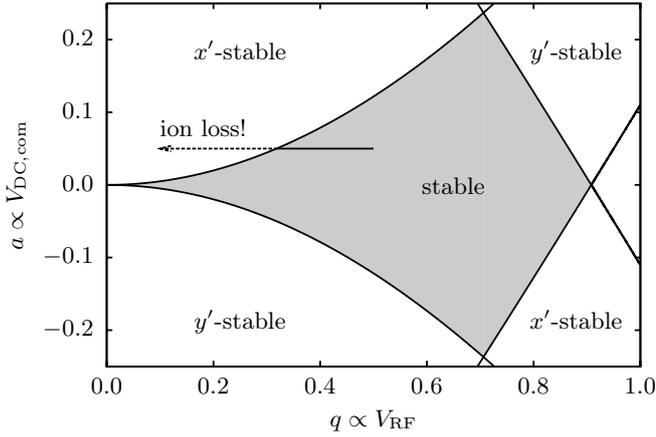

Figure 4. Lowest order stability region of Mathieu's equations of an infinitely long linear RF trap (without considering an axial confinement). The motion is stable for pairs $(q, a)$ inside the gray shaded region. Outside this region the motion is only stable in $x'$-direction or $y'$-direction or unstable in both radial directions. The arrow indicates a path in the stability diagram, when the RF voltage is decreased. It starts from a typical operational point of the RF trap. The motion of the ion will become unstable for any $a \neq 0$, before zero RF voltage is reached.

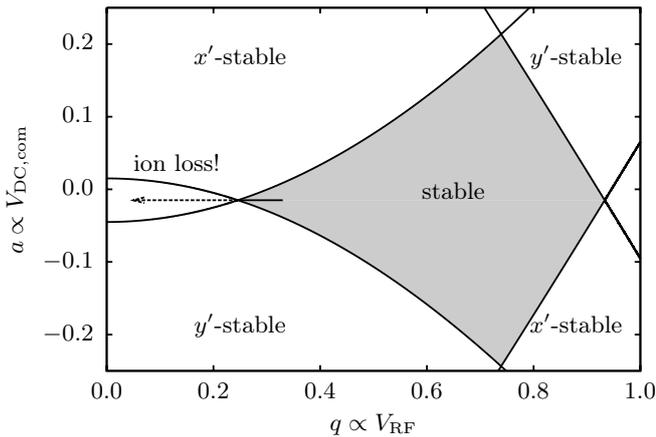

Figure 5. Example for lowest order stability region of Mathieu's equations of a linear RF trap considering a specific defocussing due to an axial confinement. We assumed some $V_{\mathrm{DC,out}} \neq 0$ and different geometry factors $\kappa_{x'} \neq \kappa_{y'}$. It demonstrates that a decrease of the RF voltage will eventually lead to an ion loss for any $a$-parameter (even $a = 0$) before zero RF voltage is reached. Additionally, it shows that $V_{\mathrm{DC,out}} = 0$ need not be the optimal value to stay inside the stable region as long as possible.

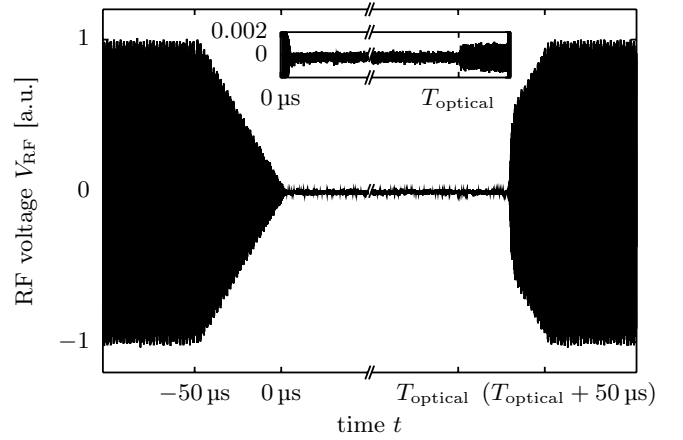

Figure 6. RF voltage of the RF trap during the experimental sequence. The amplitude is measured with a pick-up antenna inside the helical resonator, followed by filters to suppress extraneous RF signals from other sources in the laboratory. It is recorded by an oscilloscope, which is the reason for the noise. The final ramp-up of the RF voltage is slightly non-linear as the ramping speed does not meet the specifications of the RF synthesizer. The magnification (inset) shows the effect of the TTL-controlled RF-switch for times $t \leq T_{\mathrm{optical}}$. The confinement due to the residual RF voltage during $T_{\mathrm{optical}}$ can be neglected.

In general, the electrode geometry and the contact potentials due to the coating of the electrodes with Mg lead to a rotation of the $x'$ and $y'$ principal axes by an angle $\alpha$ referred to the coordinate axes and the geometry factors $\kappa_{x'/y'}$ are different. The values of $\omega_{x'}$ and $\omega_{y'}$ as well as the angle $\alpha$ are not experimentally accessible in the current setup and the contact potentials complicate predictions. A deeper analysis is presented in Sec. II C in the context of the Monte-Carlo simulation. We note that the discussion only considered an ideal quadrupole potential. Higher order contributions of the RF potential lead to instabilities within the theoretically stable regime [15], which can also result in the loss of the ion.

In the dipole trap experiment the RF amplitude is ramped down linearly in time within $50\,\mu\mathrm{s}$ for the transfer of the ion from the RF trap into the optical dipole trap and vice versa. In this way, the transfer is sufficiently slow with respect to the timescales set by any oscillation frequency. Still, instabilities are crossed comparatively fast to mitigate their impact. At the same time the transfer is slow compared to the ring-down time of the helical resonator enhancing the RF voltage for the RF trap ($1/e$ ring-down time: $0.5\,\mu\mathrm{s}$) and avoids overshoots of the oscillating circuit.

The ramp-down is achieved using the amplitude modulation (AM) mode of the RF synthesizer driving the RF

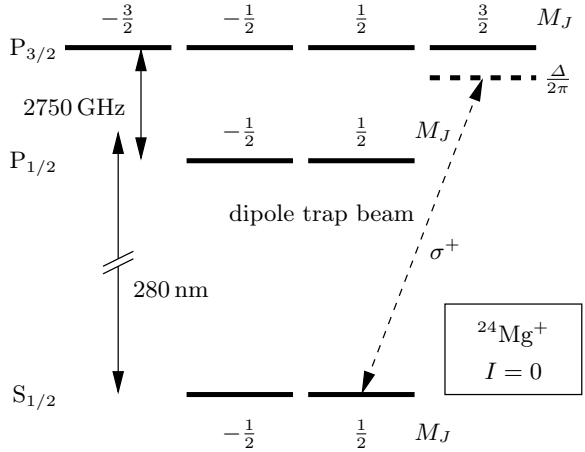

Figure 7. Relevant part of the electronic energy level scheme of $^{24}$Mg$^+$. As $^{24}$Mg$^+$ has no nuclear spin ($I = 0$), there are only one ground level (S$_{1/2}$) with angular momentum $J = 1/2$ and two exited levels (P$_{1/2}$ and P$_{3/2}$) with angular momenta $J = 1/2$ and $J = 3/2$, respectively. The $M_J$ sub-levels are shown. The dashed line indicates the $\sigma^+$-polarized dipole trap beam with a detuning of $\Delta$ from the cycling transition.

circuitry of the RF trap. The maximum attenuation by the AM amounts to approximately 99.75 %. An additional TTL-controlled RF-switch allows to further attenuate the RF voltage by a factor of $10^5$ after ramping. Hence, the total attenuation of the RF voltage during the optical trap duration $T_{\text{optical}}$, i.e. the duration of maximal attenuation of the RF potential, is on the order of $10^8$. The confinement due to the residual RF potential can therefore be neglected compared to the confinement of the dipole trap and the static electric fields. The ramping procedure is monitored on an oscilloscope using a pick-up antenna inside the helical resonator, followed by filters to suppress extraneous RF noise from other sources in the laboratory (see Fig. 6).

### D. Dipole Trap

In the following we discuss the dipole trap into which the ion is transferred during the ramp-down of the RF voltage. The laser setup for the dipole trap beam consists of a 2 W fiber laser at 1118 nm with two consecutive second harmonic generation (SHG) stages [16]. The first SHG stage has an efficiency of around 70 % and the second SHG stage of around 40 %. This leads to a power exceeding 500 mW at a wavelength of 280 nm with a highly astigmatic beam profile and a ratio of vertical to horizontal waist radii $> 10 : 1$. (At later times we even achieved a maximal power of around 600 mW.) The beam is shaped by a cylindrical and a spherical telescope. Further optical elements include an accousto-optical modulator to control its power, polarization optics, and a spherical 1 : 10 telecope with pinhole to expand and clean the mode of the beam. The expanded beam is focussed by a spherical lens with a focal length of 200 mm onto the ion. Currently, we have up to $P_{\text{optical}} = 275$ mW in the dipole trap beam.

The laser can be detuned up to $\Delta = -2\pi \times 300$ GHz red of the resonance of the S$_{1/2} \leftrightarrow$ P$_{3/2}$ transition of $^{24}$Mg$^+$ with $\omega_0 = 2\pi \times 1\,072.082\,934$ THz [17] and a natural line width of $\Gamma = 2\pi \times 41.8$ MHz [18] (see Fig. 7 for a level scheme). Thus, the maximal detuning corresponds to $\Delta \approx -7200\,\Gamma$.

During the dipole trap experiments a typical power of $P_{\text{optical}} = 190$ mW is used in the dipole trap beam and a detuning of $\Delta \approx -2\pi \times 275$ GHz. The beam is focussed on the ion and has a nearly Gaussian shape with a waist radius of $w_0 \approx 7$ μm ($1/e^2$ radius of intensity). The polarization is tuned to $\sigma^+$ (see Sec. I E). The first few off-resonantly scattered photons from the dipole trap beam pump the ion into the S$_{1/2}(M_J = 1/2)$ level. As S$_{1/2}(M_J = 1/2) \leftrightarrow$ P$_{3/2}(M_J = 3/2)$ is a cycling transition, the ion can be treated as a two-level system.

Considering solely the dipole trap potential we expect a depth of $U_0 = \frac{3c^2}{\omega_0^3} \frac{\Gamma}{\Delta} \frac{P_{\text{optical}}}{w_0^2} \approx -k_B \times 38$ mK and a maximum force perpendicular to the beam of $F_{\text{rad}} \approx 10^{-19}$ N [19]. The corresponding trapping frequencies at the approximately harmonic bottom of the dipole potential amount to $\omega_{\text{rad}} = \sqrt{-4U_0/(mw_0^2)} \approx 2\pi \times 165$ kHz in the radial directions. The confinement in the direction of beam propagation is $\omega_k \approx 2\pi \times 2$ kHz and can be neglected.

We note that, on the one hand, the beam profile of the dipole trap beam is not perfectly round due to experimental imperfections, but has waist radii of $w_{0,h} \approx 5$ μm in the horizontal direction ($x$–$z$–plane) and $w_{0,v} \approx 7$ μm in the vertical $y$-direction. This is a relict of the beam shape from the second SHG stage and can hardly be compensated without sacrificing too much laser power. On the other hand, any deviation from a perfect Gaussian mode (due to, e.g. abberations, diffraction, . . . ) easily leads to a reduction of the peak intensity and thus a reduction of the depth of the dipole trap potential. We think that the trap depth calculated for the round beam with $w_0 = 7$ μm with an measurement error (including both statistical and systematic errors) of 1 μm remains an appropriate estimate.

### E. Optimization of the Polarization

Besides of the shape of the dipole trap beam, a pure polarization is important to allow for the maximal dipole trap depth and minimal heating (see the following Sec. I F). The polarization is adjusted to $\sigma^+$ referred to a quantization axis provided by a magnetic field of $B = 5.5$ G. The magnetic field is parallel to the propagation direction of the dipole trap beam. The polarization is optimized using a $^{25}$Mg$^+$ ion that has a nuclear spin of $I = 5/2$. Hence its ground state is hyperfine-split into the S$_{1/2}(F = 2)$ and S$_{1/2}(F = 3)$ manifolds with an energy difference of $2\pi\hbar \times 1.8\,\mathrm{GHz}$. Neigbouring $M_F$ sub-levels are Zeeman-shifted by $\approx 2\pi \times 3\,\mathrm{MHz}$ by the magnetic field. The dipole trap beam at a low intensity is tuned close to the $\mathrm{S}_{1/2}(F=3, M_F=3) \leftrightarrow \mathrm{P}_{3/2}(F=4, M_F=4)$ cycling transition. Polarization components other than $\sigma^+$ can lead the ion outside the cycling transition into the $\mathrm{S}_{1/2}(F=2)$ ground state, which is off-resonant. Hence, the ion appears "dark" in state $\mathrm{S}_{1/2}(F=2)$ and fluoresces "bright" in state $\mathrm{S}_{1/2}(F=3)$. The optimization of the polarization is performed by analysing the fluorescence of the ion and adjusting polarization optics accordingly. Small coils producing magnetic fields perpendicular to the main field are used align the quantization axis with the direction of beam propagation. This yields a $\sigma^+$ purity on the order of typically $10^5 : 1$ (intensity). In the experiments with $^{24}\mathrm{Mg}^+$ the polarization is not re-adjusted on short timescales, still allowing for a modest estimate for the purity of better than $1000 : 1$.

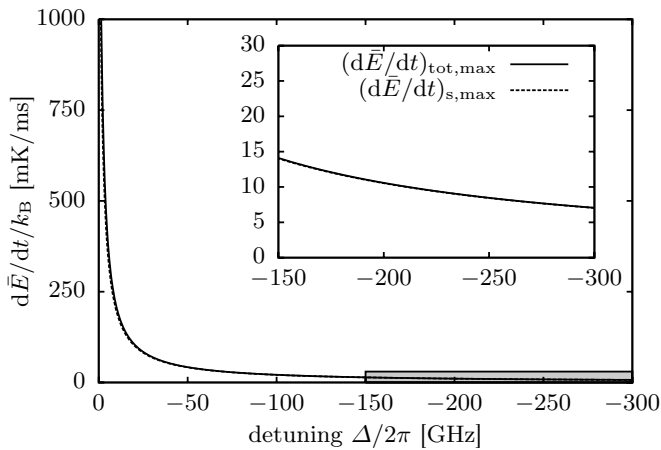

Figure 8. Theoretically expected maximal total energy gain $(\mathrm{d}\bar{E}/\mathrm{d}t)_{\mathrm{tot,max}}$ and maximal energy gain due to recoil heating $(\mathrm{d}\bar{E}/\mathrm{d}t)_{\mathrm{s,max}}$ as a function of the detuning $\Delta$ for a Gaussian beam dipole trap (according to Ref. 20). The graphs are calculated for a waist radius $w_0 = 7\,\mu\mathrm{m}$, a potential depth $U_0 = -k_\mathrm{B} \times 38\,\mathrm{mK}$, an position in direction of the beam propagation $z_k = 0$ of the ion, and the radial position $r$ with maximal heating rate. The gray shaded area is magnified in the inset, which allows to extract the energy gain for the experimental parameters. The two curves are almost identical such that for the given parameters recoil heating is expected to be dominant.

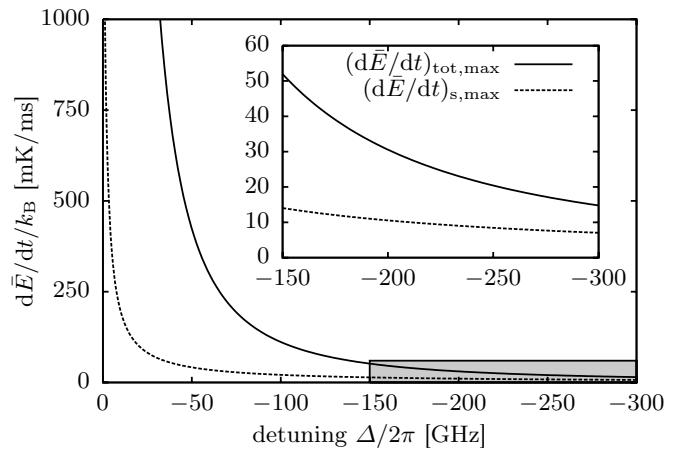

Figure 9. Theoretically expected maximal total energy gain $(\mathrm{d}\bar{E}/\mathrm{d}t)_{\mathrm{tot,max}}$ and maximal energy gain due to recoil heating $(\mathrm{d}\bar{E}/\mathrm{d}t)_{\mathrm{s,max}}$ as a function of the detuning $\Delta$ for a standing-wave dipole trap created by retro-reflecting the Gaussian beam (according to Ref. 20). The graphs are calculated for a waist radius $w_0 = 7\,\mu\mathrm{m}$, a potential depth $U_0 = -k_\mathrm{B} \times 38\,\mathrm{mK}$, a radial position $r = 0$ of the ion, and the position in direction of the beam propagation $z_k$ with maximal heating rate. The power $P_{\mathrm{optical}}(\Delta)$ of the dipole trap beam corresponds to a quarter of the power of Fig. 8 to account for the retro-reflection and interference in the standing wave. The gray shaded area is magnified in the inset. The total heating rate approximately doubles at the detunings used in the single-beam dipole trap experiments.

The Zeeman shift of the $M_J$-levels in $^{24}\mathrm{Mg}^+$ amounts to approximately $\pm 2\pi \times 9\,\mathrm{MHz}$, which is negligible compared to the AC Stark shift due to the dipole trap beam on the order of $2\pi \times 800\,\mathrm{MHz}$.

### F. Heating Effects in Dipole Traps

The analysis of the experimental data requires the knowledge about the relevant heating effects in dipole traps. Heating effects in dipole traps have been theoretically examined in the 1980s in Refs. 20 and 21 for a two-level system with negligible velocity. The main heating effects were determined to be recoil heating and heating due to dipole force fluctuations. They are quantified in terms of a momentum diffusion constant $2D_p$, which results in an average energy gain of $\mathrm{d}\bar{E}/\mathrm{d}t = D_p/m$ for the trapped particle.

The recoil heating is caused by a momentum diffusion due to absorption of laser photons and the subsequent spontaneous emissions resulting in a random walk in momentum space. Each of the absorptions and spontaneous emissions increases the expectation value of the variance of the momentum $\bar{\sigma}_p^2$ by $(\hbar k)^2$, where $k = \omega_0/c$ denotes the wavenumber of the laser photons. Thus, the energy gain can be written as $(\mathrm{d}\bar{E}/\mathrm{d}t)_\mathrm{s} = \Gamma_\mathrm{s} \times 2E_\mathrm{rec}$ with the scattering rate $\Gamma_\mathrm{s}$ and recoil energy $E_\mathrm{rec} = (\hbar k)^2/(2m) \approx k_\mathrm{B} \times 5\,\mu\mathrm{K}$. The recoil heating rate has its maximum $\Gamma_\mathrm{s,max} = (U_0/\hbar)(\Gamma/\Delta)$ at the center of the dipole potential, where the intensity has its maximum. For a potential depth of $U_0 = -k_\mathrm{B} \times 38\,\mathrm{mK}$, a detuning of $\Delta = -2\pi \times 275\,\mathrm{GHz}$, and a waist radius of $w_0 = 7\,\mu\mathrm{m}$ we obtain $\Gamma_\mathrm{s,max} \approx 750\,\mathrm{ms}^{-1}$ and $(\mathrm{d}\bar{E}/\mathrm{d}t)_\mathrm{s,max} \approx k_\mathrm{B} \times 7.7\,\mathrm{mK/ms}$.

Dipole force fluctuations are due to the opposite signs of the dipole forces for an ion in the states $\mathrm{S}_{1/2}(M_J = 1/2)$ and $\mathrm{P}_{3/2}(M_J = 3/2)$. In the dressed-state approach of Ref. 21 spontaneous emissions between levels subject to different dipole forces lead to a momentum diffusion and energy gain, respectively. The diffusion constant is proportional to the square of the intensity gradient. Therefore, the energy gain due to dipole force fluctuations is highest close to the gradients of the dipole potential.

The full mathematical treatment for a two-level atom at rest can be found in Ref. 20. In Fig. 8 the total energy gain $(\mathrm{d}\bar{E}/\mathrm{d}t)_\mathrm{tot}$ as a function of the detuning $\Delta$ is compared to the energy gain solely due to recoil heating $(\mathrm{d}\bar{E}/\mathrm{d}t)_\mathrm{s}$. The potential depth $U_0$ is kept constant by adapting the power $P_\mathrm{optical}(\Delta)$ as a function of the detuning. The difference $(\mathrm{d}\bar{E}/\mathrm{d}t)_\mathrm{tot} - (\mathrm{d}\bar{E}/\mathrm{d}t)_\mathrm{s}$ amounts to only $k_\mathrm{B} \times 0.02\,\mathrm{mK/ms}$ at $\Delta = -2\pi \times 275\,\mathrm{GHz}$. Thus, at the typical experimental parameters recoil heating is expected to be dominant.

However, for a standing wave dipole trap created by retro-reflecting the Gaussian beam, the heating due to dipole force fluctuations has to be considered (see Fig. 9). This is due to the comparatively large intensity gradients in direction of the standing wave (the potential now has a modulation on length scales $\lambda/2 \ll w_0$). The power $P_\mathrm{optical}(\Delta)$ was reduced by a factor of 4 compared to Fig. 8 to maintain the potential depth of $U_0 = -k_\mathrm{B} \times 38\,\mathrm{mK}$. The maximal total energy gain increases by roughly a factor of two at our experimental parameters compared to the trap consisting of a single Gaussian beam: $(\mathrm{d}\bar{E}/\mathrm{d}t)_\mathrm{tot,max} \sim k_\mathrm{B} \times 15\,\mathrm{mK/ms}$.

The dependence of the heating rate due to dipole force fluctuation on the Rabi frequency $\Omega_\mathrm{R}$ can be estimated from Eq. (4.6) of Ref. 21. It yields $D_p \propto \Omega_\mathrm{R}^8$ for detunings $\Delta \gg \Omega_\mathrm{R}$. As $U_0 \propto \Omega_\mathrm{R}^2$, we expect an increase of the heating rate by a factor of 16 for a doubling of the potential depth at $\Delta = -2\pi \times 275\,\mathrm{GHz}$. The exact value according to Ref. 20 is $(\mathrm{d}\bar{E}/\mathrm{d}t)_\mathrm{tot,max} \approx k_\mathrm{B} \times 220\,\mathrm{mK/ms}$, which emphasizes the strong dependence on the intensity. This adds difficulty to dipole trap experiments with a standing wave.

Further heating in the dipole trap can be caused by an impure polarization. The coupling of the dipole trap beam with $\sigma^+$ polarization to the $S_{1/2}(M_J = -1/2) \leftrightarrow P_{3/2}(M_J = 1/2)$ transition is a factor of $1/\sqrt{3}$ weaker than to $S_{1/2}(M_J = 1/2) \leftrightarrow P_{3/2}(M_J = 3/2)$. Polarization components other than $\sigma^+$ can scatter the ion into the $S_{1/2}(M_J = -1/2)$ level. A fluctuation between the $S_{1/2}(M_J = 1/2)$ and $S_{1/2}(M_J = -1/2)$ levels leads to force fluctuations and thus a momentum diffusion (see, e.g., [22]), which is similar to the diffusion caused by dipole force fluctuations above. Additionally, the potential depth in the $S_{1/2}(M_J = -1/2)$ level is reduced to $U_0/3$, which can lead to an earlier loss of the ion. However, we expect this effect to be negligible in our experiment (see Sec. I E). Heating due to intensity noise in the dipole trap beam [23] is also estimated to be orders of magnitude smaller than the recoil heating rate.

## II. EXPERIMENTAL RESULTS AND ANALYSIS

The aim of the measurements is to obtain an estimate for the trapping probability $P$ of an ion in the dipole trap for different parameters $T_\mathrm{optical}$ and $P_\mathrm{optical}$. $P$ is determined as the mean number of successful trapping attempts with approximately 40 ions for each set of parameters. For example, this corresponds to more than 200 trapping attempts according to the above scheme for $T_\mathrm{optical} = 1\,\mathrm{ms}$ and $P_\mathrm{optical} = 190\,\mathrm{mW}$ (cmp. Fig. 10).

Results of the dependence of the trapping probability on $T_\mathrm{optical}$ and $P_\mathrm{optical}$ have been published in Ref. 1. To summarize, the half-life in the dipole trap can be estimated to $T_\mathrm{optical}^{(1/2)} = 2.4\,\mathrm{ms}$ and to be mainly limited by the intrinsic recoil heating by the dipole trap beam. Furthermore, a measurement with $P_\mathrm{optical} = 0\,\mathrm{mW}$ resulting in zero trapping probability confirms that there is no other relevant three-dimensional trapping potential during $T_\mathrm{optical}$.

### A. Analytic Models

The experimental data presented in Ref. 1 have been analysed using a simple analytic model. In this section, the analytic model (cmp. Ref. 1) is derived, and the impact of the simplifications is discussed. Based on the conclusions an improved analytic model is developed.

For the first analytic model the total potential consisting of dipole trap and static electric potential is approximated as a spherically symmetric harmonic potential with frequency $\omega_\mathrm{harm}$:

$$U_\mathrm{harm}(\vec{r}) = U_0 + \frac{1}{2}m\omega_\mathrm{harm}^2 r^2. \quad (4)$$

(Note that $U_0 < 0$ by definition.) In the model the particle energies $E$ are subject to the Boltzmann distribution of this potential:

$$f(E) = \frac{1}{2}\beta^3(E-U_0)^2 \mathrm{e}^{-\beta(E-U_0)} \quad \text{with} \quad \beta = \frac{1}{k_\mathrm{B}T}, \quad (5)$$

where $T$ denotes the temperature.

A particle will be considered as trapped, if it has a total energy $E \leq 0$ after the dipole trap duration $T_\mathrm{optical}$. That is why the model is referred to as the *energy-cutoff* model in the following. The trapping probability for a particle is obtained as [24]

$$P_\mathrm{en}(\xi) = \int_{U_0}^{0} \mathrm{d}E\, f(E) = 1 - \left(\frac{\xi^2}{2} + \xi + 1\right)\mathrm{e}^{-\xi} \quad (6)$$

with $\xi := \beta|U_0|$.

To model the dipole trap experiments, the temperature is assumed to increase linearly in time $t$ with heating rate $R := \mathrm{d}T/\mathrm{d}t$ from an initial temperature $T_0$: $T(t) = T_0 + Rt$. Both the potential depth $|U_0| =: |u_0|\,P_\mathrm{optical}$ and the heating rate $R =: \tilde{R}\,P_\mathrm{optical}$ (see Sec. I F) are

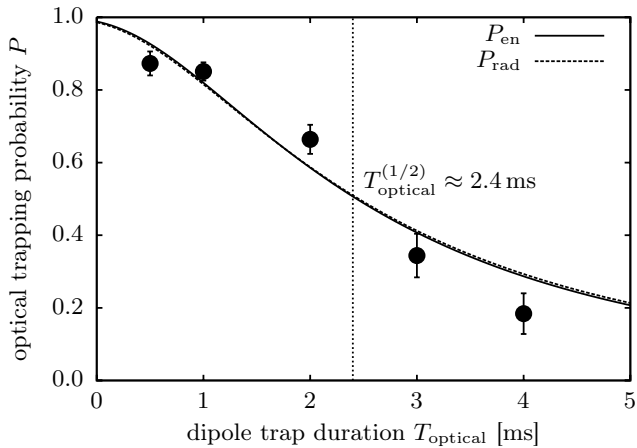

Figure 10. Comparison of the energy-cutoff and radial-cutoff model: The optical trapping probabilities $P_\text{en}$ according to the energy-cutoff model and $P_\text{rad}$ according to the radial-cutoff model are shown as functions of the dipole trap duration $T_\text{optical}$. The points represent the measured optical trapping probabilities and the error bars the standard deviation of the mean [1]. The parameters for which the curves of the analytic models are shown originate from a joint fit to the experimental data of this figure and Fig. 11. While the curves are almost identical, the best fit values for the initial temperature and the heating rate are approximately 20 % greater for the radial-cutoff model. Experimental parameters: $w_0 = (7 \pm 1)\,\mu\text{m}$, $\Delta = -2\pi \times 275\,\text{GHz}$, $P_\text{optical} = 190\,\text{mW}$, and $\omega_z = 2\pi \times 47\,\text{kHz}$.

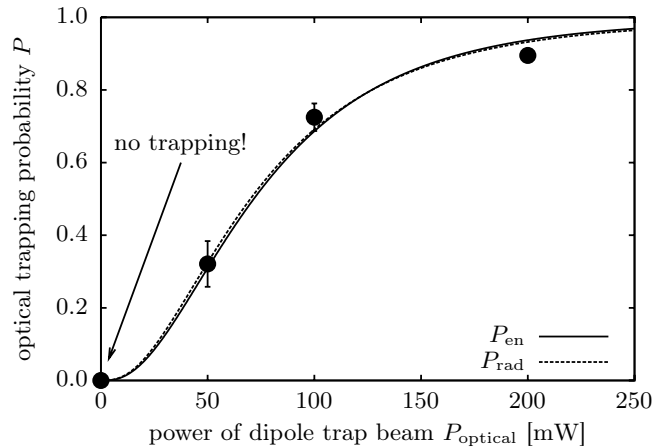

Figure 11. Comparison of the energy-cutoff and radial-cutoff model (analogously to Fig. 10): The optical trapping probabilities $P_\text{en}$ and $P_\text{rad}$ according to the models are shown as functions of the power $P_\text{optical}$ of the dipole trap beam. Experimental parameters: $w_0 = (6.5 \pm 1.0)\,\mu\text{m}$, $\Delta = -2\pi \times 300\,\text{GHz}$, $T_\text{optical} = 0.5\,\text{ms}$, and $\omega_z = 2\pi \times 41\,\text{kHz}$.

assumed to be proportional to the power of the dipole trap beam. Hence, we obtain

$$\xi = \frac{|u_0| P_\text{optical}}{k_\text{B}(T_0 + \tilde{R} P_\text{optical}\, t)}. \tag{7}$$

In the experiments the heating in the direction of beam propagation is larger, because the absorption of photons always occurs in this direction as opposed to the emission of photons. However, the total potential in the dipole trap experiments is anharmonic (see also Secs. II B and II C), such that we expect cross-dimensional mixing. This justifies the assumption of an isotropic heating in the model.

Next, we substitute $\xi$ in Eq. (6) with Eq. (7) and use the predicted value $|u_0| = k_\text{B} \times 38\,\text{mK}/190\,\text{mW}$. A joint fit of $P_\text{en}$ for the data of Figs. 10 and 11 yields parameters $T_0 = 4.6\,\text{mK}$ and $R = 3.9\,\text{mK/ms} \times P_\text{optical}/190\,\text{mW}$. Here, the uncertainty of the waist of $\pm 1\,\mu\text{m}$ generates an error of $|u_0|$ and, hence, of the fitted parameters of 30 %. The time-derivative of the relation $\bar{E} = 3k_\text{B}T + U_0$ which holds for a three-dimensional harmonic potential yields $\text{d}\bar{E}/\text{d}t = 3k_\text{B}R$. Hence, the value of the heating rate from the fit yields an energy gain $\text{d}\bar{E}/\text{d}t = k_\text{B} \times 11.7\,\text{mK/ms}$, which is a factor of 1.5 higher than the estimated recoil heating rate at the center of the dipole trap (cmp. Sec. I F) [25]. The following discussion will show that the experimental results are still in agreement with recoil heating as the relevant heating effect predicted in Sec. I F.

Now we further analyse the energy-cutoff model. A major simplification is the cutoff of the probability distribution in energy space. However, in the experiment the cutoff happens in position space or equivalently in *potential* energy space. Additional energy of up to $|U_0|$ can be stored in the angular momentum of a trapped particle. The energy-cutoff model is extended to a *radial-cutoff* model taking this into account by following a calculation in Ref. 26. The probability for a particle to be trapped at time $t$ is defined as the probability to have an orbit with potential energy below zero. This is still an approximation, because a small fraction of the particles is on orbits with potential energy larger than zero at intermediate times, but evolves back to lower orbits at a later time. These particles are not treated as lost in the radial-cutoff model. The possibility to evolve back to lower orbits arises from the diffusive nature of the heating process.

In spherical coordinates the energy of the ion is given by

$$E = \frac{1}{2m}\left(p_r^2 + \frac{L^2}{r^2}\right) + U_\text{harm}(\vec{r}), \tag{8}$$

where $p_r = m\,\text{d}r/\text{d}t$ denotes the momentum in radial direction and $L$ the angular momentum. The maximal radius of an orbit is obtained for $p_r = 0$:

$$r_\text{max}(E, L) = \left(\frac{(E - U_0) + \sqrt{(E - U_0)^2 - (\omega_\text{harm} L)^2}}{m\omega_\text{harm}^2}\right)^{1/2}. \tag{9}$$

We define a characteristic function

$$\Theta(E, L) = \begin{cases} 1 & \text{if } U_{\text{harm}}(r_{\text{max}}) \leq 0 \\ 0 & \text{else} \end{cases}, \quad (10)$$

which is 1 only if the potential energy of an orbit does not exceed zero. For a harmonic potential the characteristic function can be rewritten as

$$\Theta(E, L) = \begin{cases} 1 & \text{if } U_0 \leq E \leq 0 \\ 1 & \text{if } 0 < E \leq -U_0 \text{ and } L \geq \frac{\sqrt{-U_0 E}}{\omega_{\text{harm}}/2} \\ 0 & \text{else} \end{cases}. \quad (11)$$

Hence, if a particle has an energy $U_0 \leq E \leq 0$, it will always be trapped in the potential. If it has an energy $0 < E \leq -U_0$, it may (but need not) be trapped, because in addition to the energy of up to $|U_0|$ that can be stored as potential/radial kinetic energy, there is also kinetic energy of up to $|U_0|$ related to the angular momentum.

With some effort, the trapping probability can be calculated analytically as a phase space integral:

$$P_{\text{rad}}(\xi) = \int d\Omega\, W(E)\, \Theta(E, L) = 1 - e^{-2\xi} - 2\xi e^{-\xi} \quad (12)$$

with the infinitesimal phase space volume $d\Omega = d^3r d^3p$ and the density of states per phase space volume

$$W(E) = \left(\frac{\omega_{\text{harm}}\beta}{2\pi}\right)^3 e^{-\beta(E - U_0)}. \quad (13)$$

The comparison of the trapping probabilities of the two analytical models shows $P_{\text{rad}}(\xi) \approx P_{\text{en}}(1.2\,\xi)$.

Again we substitute $\xi$ in Eq. (12) with Eq. (7) and fit $P_{\text{rad}}$ to the measured data analogously to the fit of the energy-cutoff model. The best fit results for both the heating rate as well as the initial temperature are approximately 20 % higher than for the energy-cutoff model, which could be expected due to $P_{\text{rad}}(\xi) \approx P_{\text{en}}(1.2\,\xi)$. This shows that energy stored as angular momentum is of limited relevance. The fitted curves corresponding to both models are compared in Figs. 10 and 11 and are almost identical.

To understand the deviations between experimental results and analytic models, we have to further investigate the simplifications in the latters: Both models simplify the total potential of the trap as an isotropic harmonic potential. In contrast, the total potential in the dipole trap experiments is asymmetric and its well depth is decreased in some directions due to the defocusing effect of the static electric potential. However, both analytic models still agree with the experimental results within a factor of two and are useful for rough estimates of expected trapping probabilities. The potential shape and well depth will be further discussed in Sec. II B and the Monte-Carlo simulation presented in Sec. II C will finally overcome most simplifications of the analytic models.

Before we conclude this section, we extended the calculations of the radial-cuttoff model to analyse the scattering rate. In the dipole trap potential it is proportional to the intensity and thus the potential itself: $\Gamma_s(\vec{r}) \propto -U(\vec{r})$. For the harmonic potential in Eq. (4), the corresponding scattering rate becomes

$$\Gamma_{s,\text{harm}}(\vec{r}) = \begin{cases} \Gamma_{s,\text{max}} \times U_{\text{harm}}(\vec{r})/U_0 & \text{for } U_{\text{harm}}(\vec{r}) \leq 0 \\ 0 & \text{else} \end{cases}. \quad (14)$$

The average scattering rate can be calculated analogously to $P_{\text{rad}}(\xi)$:

$$\bar{\Gamma}_{s,\text{harm}}(\xi) = \frac{1}{P_{\text{rad}}(\xi)} \int d\Omega\, W(E)\, \Theta(E, L)\, \Gamma_{s,\text{harm}}(\vec{r}) \quad (15)$$

$$= \Gamma_{s,\text{max}} \left(1 - \frac{3 - (2\xi + 3)e^{-2\xi} - 2\xi(\xi + 2)e^{-\xi}}{2\xi(1 - e^{-2\xi} - 2\xi e^{-\xi})}\right). \quad (16)$$

For zero temperature this yields $\bar{\Gamma}_{s,\text{harm}} = \Gamma_{s,\text{max}}$. In the limit $k_B T \to \infty$ the avarage scattering rate decreases to $\bar{\Gamma}_{s,\text{harm}} \to \Gamma_{s,\text{max}}/2$. At $k_B T = |U_0|/3$ (corresponding to $\bar{E} = 0$), it reaches $\bar{\Gamma}_{s,\text{harm}} = 0.65\Gamma_{s,\text{max}}$. This result shows that the avarage scattering rate is always greater than $\Gamma_{s,\text{max}}/2$ (as opposed to the rough estimate in Ref. 1). A simplification in this calculation still is the isotropic decrease of the scattering rate in all three dimensions, while in the experiment it remains constant along the propagation direction of the dipole trap beam to a good approximation.

### B. Analysis of Potential Shape and Well Depth

The potential of the dipole trap close to the waist of the laser beam (distance to waist small compared to the Rayleigh length $z_0$) can be approximated by

$$U_{\text{optical}}(\vec{r}) \approx U_0 \exp\left(-\frac{2}{w_0^2}\left(\frac{(x-z)^2}{2} + y^2\right)\right), \quad (17)$$

where the dipole trap beam propagates in direction $\hat{k} = (1, 0, 1)/\sqrt{2}$. The trapping frequency at the approximately harmonic bottom of this potential is $\omega_{\text{rad}} = \sqrt{-4U_0/(mw_0^2)}$. To simplify matters, we assume that the principal axes of the static electric potential $U_{\text{DC}}$ coincide with the coordinate axes (i.e., the angle between principal axes and coordinate axes is $\alpha = 0\,°$):

$$U_{\text{DC}}(\vec{r}) = \frac{1}{2}m\left(\omega_x^2 x^2 + \omega_y^2 y^2 + \omega_z^2 z^2\right). \quad (18)$$

The total potential is the superposition of these two potentials: $U(\vec{r}) = U_{\text{optical}}(\vec{r}) + U_{\text{DC}}(\vec{r})$.

The well depth of $U(\vec{r})$ differs for different directions, because $U_{\text{DC}}(\vec{r})$ has a defocusing effect perpendicular to the $z$-axis. At some points the potential $U(\vec{r})$ turns from focussing to defocussing for values $U(\vec{r}) < 0$, i.e., the well depth is reduced compared to the depth $|U_0|$ of the dipole trap potential only. We are interested in the minimal particle energy $U_{\text{mpe}}$ which is neccessary to escape from the trap. (The minimal well depth can then be written as $|U_0 - U_{\mathrm{mpe}}|$.) Any particle with an energy $E < U_{\mathrm{mpe}}$ will be trapped independently of its trajectory. A particle with larger energy may be trapped for a certain duration depending on its trajectory.

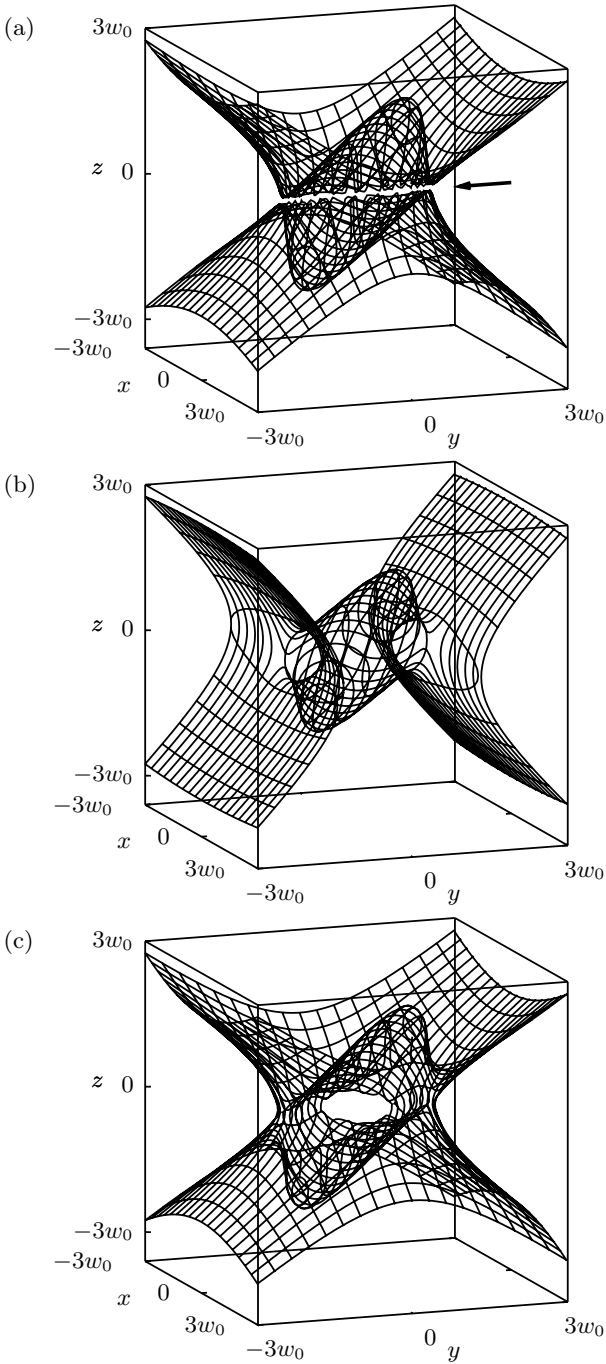

Figure 12. Equipotential surfaces $U(\vec{r}) = U_{\mathrm{mpe,max}}$ of the total potential. The parameters of the graphs correspond to the experimental parameters $U_0 = -k_B \times 38\,\mathrm{mK}$, $\omega_z = 2\pi \times 47\,\mathrm{kHz}$, and $w_0 = 7\,\mathrm{\mu m}$. The frequency in $y$-direction is (a) $\omega_y^2 = \omega_{y,\mathrm{max}}^2$, (b) $\omega_y^2 = (1.2\,\omega_{y,\mathrm{max}})^2$, and (c) $\omega_y^2 = (0.95\,\omega_{y,\mathrm{max}})^2$, respectively (see text). The points defining $U_{\mathrm{mpe,max}}$ in (a) appear as "cut" trough the egg-like surface enclosing the trap centre (indicated by the arrow). Note that $\omega_{y,\mathrm{max}}^2$ is negative, i.e., an increase of the absolute value of $\omega_y^2$ leads to a stronger defocussing effect in $y$-direction (b), while a decrease leads to a stronger defocussing in a direction in the $x$-$z$-plane (c).

To calculate $U_{\mathrm{mpe}}$ we derive the critical points $\vec{r}_{\mathrm{crit}}$ of the potential, which fulfill $\nabla U(\vec{r})|_{\vec{r}_{\mathrm{crit}}} = 0$. By symmetry there are at least two such points $\pm\vec{r}_{\mathrm{mpe}}$ with $U(\pm\vec{r}_{\mathrm{mpe}}) = U_{\mathrm{mpe}}$ in addition to the local minimum at $\vec{r}_0 = (0,0,0)$ with $U(\vec{r}_0) = U_0$ (as long as the potential provides a three-dimensional confinement). We give solutions for some special cases in the following.

Except for $\omega_x$ and $\omega_y$ all parameters of the potential are sufficiently well known in the experiments (cmp. Sec. I). Therefore, we analyse $U_{\mathrm{mpe}}$ as a function of $\omega_y$ in the following, while $\omega_x = [-(\omega_y^2 + \omega_z^2)]^{1/2}$ is varied along with $\omega_y$ to meet Laplace's equation. An interesting case is the maximum of $U_{\mathrm{mpe}}$, because it sets the deepest trap we can expect. It is obtained for $\omega_{y,\mathrm{max}}^2 = (1-\sqrt{3})\,\omega_z^2$ and amounts to $U_{\mathrm{mpe,max}} = b\,U_0$, with

$$b = p\,[1 - \ln(p)] \qquad \text{and} \qquad p = \frac{(\sqrt{3}-1)\,\omega_z^2}{\omega_{\mathrm{rad}}^2}. \qquad (19)$$

By inserting $\omega_z = 2\pi \times 47\,\mathrm{kHz}$ and $\omega_{\mathrm{rad}} = 2\pi \times 165\,\mathrm{kHz}$ for the experimental parameters (cmp. Fig. 10) we obtain a defocussing frequency in $y$-direction of $\omega_{y,\mathrm{max}}^2 = -(2\pi \times 40\,\mathrm{kHz})^2$, $b \approx 0.23$, and $U_{\mathrm{mpe,max}} = -k_B \times 8.6\,\mathrm{mK}$. This means that particles see a minimal well depth amounting to only $|U_0 - U_{\mathrm{mpe}}| = k_B \times 29.4\,\mathrm{mK}$ in some directions even for an "optimal" choice of $\omega_y$. The well depth is reduced by already a quarter compared to $|U_0| = k_B \times 38\,\mathrm{mK}$.

The potential shape for $\omega_y = \omega_{y,\mathrm{max}}$ is special insofar as there is an infinite number of critical points $\vec{r}_{\mathrm{mpe}}$ lying on a "ring". This is graphically illustrated in Fig. 12a, where the equipotential surface for $U(\vec{r}) = U_{\mathrm{mpe,max}}$ shows these points as "cut" (indicated by an arrow) through the egg-like surface containing the trap centre. A particle with energy $E = U_{\mathrm{mpe,max}}$ can only escape from the trap through this "cut". For a larger defocussing in $y$-direction we obtain $U_{\mathrm{mpe}} < U_{\mathrm{mpe,max}}$. Two "holes" in the equipotential surface $U(\vec{r}) = U_{\mathrm{mpe,max}}$ form on the $y$-axis (see Fig. 12b). Analogously, a larger defocussing in $x$-direction leads to two "holes" in the $x$-$z$-plane (see Fig. 12c). In the cases of Figs. 12b and 12c, a particle with energy $E < U_{\mathrm{mpe,max}}$ can escape from the trap in direction of the "holes".

Another interesting case is the value of $\omega_y$, where the total potential ceases to provide a three-dimensional confinement. Two cases have to be distinguished: Firstly, the defocussing in $y$-direction trivially becomes larger than the focussing effect of the dipole trap beam for $\omega_y^2 = -\omega_{\mathrm{rad}}^2 \approx -(2\pi \times 165\,\mathrm{kHz})^2$ (again for the experimental parameters; cmp. Fig. 10). Secondly, for the $x$-direction the total potential becomes repelling for a (nontrivial) value of $\omega_x^2 \approx -(2\pi \times 43.5\,\mathrm{kHz})^2$ corresponding to $\omega_y^2 \approx -(2\pi \times 17.8\,\mathrm{kHz})^2$. The higher sensitivity to a defo-cussing in the $x$-direction can be explained by the angle of $45\,°$ between the dipole trap beam and the $x$-axis.

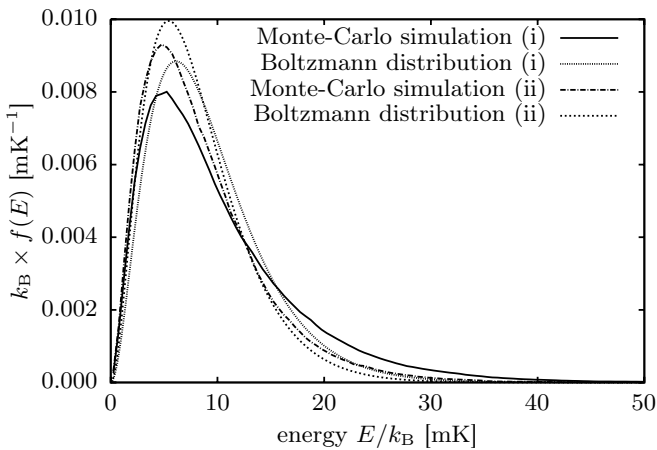

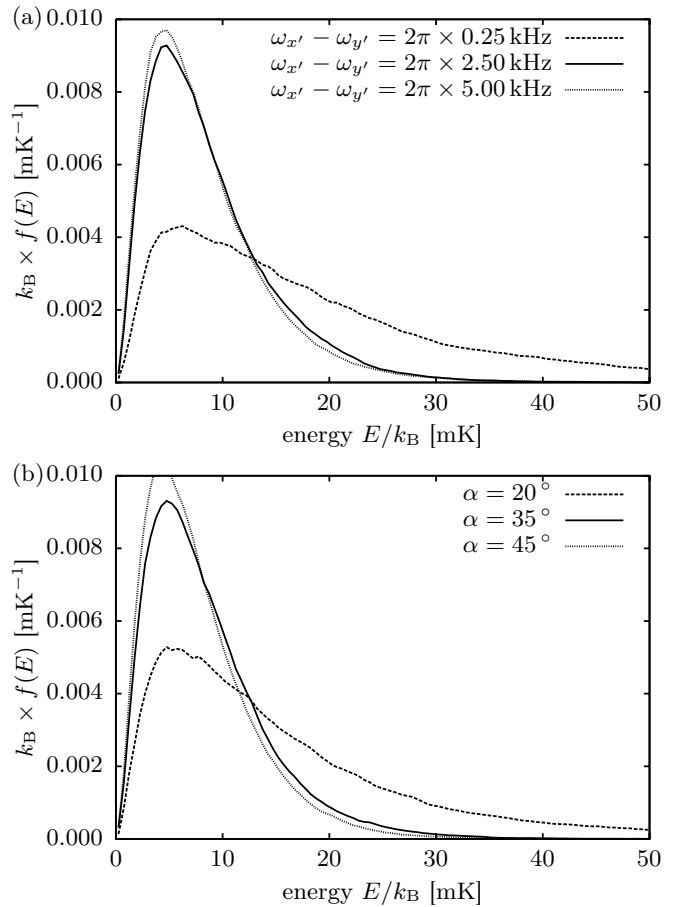

Figure 13. Probability distributions $f(E)$ of the particle energies $E$ during Doppler cooling in the RF trap. The curves show the results from the Monte-Carlo simulation and related fits of the Boltzmann distribution of the three-dimensional harmonic potential. The frequency splitting is set to $\omega_{x'} - \omega_{y'} = 2\pi \times 2.5\,\text{kHz}$ and the tilting angle to $\alpha = 35\,°$. The simulation is performed for a cooling duration of 2 s with time steps of $\Delta t = 0.1/\Gamma$. "(i)" denotes the parameters used during Doppler cooling corresponding to Fig. 10: $\omega_z = 2\pi \times 47\,\text{kHz}$, $\omega_{x'} = 2\pi \times 950\,\text{kHz}$, $P_{\text{DL}} = 8.6\,\mu\text{W}$, and $P_{\text{DLdet}} = 270\,\mu\text{W}$. The axial displacement of the ion due to the light pressure of the Doppler cooling beams amounts to $\Delta z = 8.8\,\mu\text{m}$. The radial displacement is of the order of few nanometres and can be neglected. The energies $E$ are referred to the harmonic oscillator shifted by $\Delta z$. The curves "(ii)" correspond to Fig. 11: $\omega_z = 2\pi \times 42\,\text{kHz}$, $\omega_{x'} = 2\pi \times 880\,\text{kHz}$, $P_{\text{DL}} = 2.1\,\mu\text{W}$, and $P_{\text{DLdet}} = 138\,\mu\text{W}$. These parameters yield a displacement due to light pressure of $\Delta z = 2.8\,\mu\text{m}$. The curves of the corresponding Boltzmann distributions fit well to the simulated curves. The temperatures according to the fits of the Boltzmann distribution are $T^{(\text{i})} = 3.1\,\text{mK}$ for the parameters of (i) and $T^{(\text{ii})} = 2.7\,\text{mK}$ for the parameters of (ii).

Figure 14. Comparison of the probability distributions $f(E)$ of the particle energies $E$ during Doppler cooling in the RF trap for (a) different splittings $\omega_{x'} - \omega_{y'}$ and (b) angles $\alpha$. The curves show the results from the Monte-Carlo simulation with the same parameters as used in Fig. 13 for parameters (ii). For small frequency splitting $\omega_{x'} - \omega_{y'}$ or small angles $\alpha$ Doppler cooling becomes inefficient.

The discussion reveals that the absolute value of the potential depth assumed in the analytic models (see Sec. II A) is overestimated, while the scattering rate is still determined by the laser intensity in the dipole trap beam. However, a certain $U_{\text{mpe}}$ does not imply that particles with larger energy are eventually lost from the trap within $T_{\text{optical}}$. Therefore, the analysis of the impact of the anisotropy of the potential requires a more realistic model, which is given in the following Sec. II C.

### C. Monte-Carlo Simulation

In order to overcome the limitations of the models from Sec. II A, a rate equation Monte-Carlo simulation has been developed. Therein, the ion has a classical position $\vec{r}$ and velocity $\vec{v}$ and a discrete state $i$ out of the level scheme depicted in Fig. 7. The simulation consists of two parts: Firstly, the Doppler cooling process in the RF trap is simulated. This simulation does not only allow to analyse the impact of the static electric potential on the efficiency of the Doppler cooling process in the RF trap. It also provides the initial values (position, velocity, internal state) of the ions for the simulation of the dipole trap experiments and thus no free parameter for the initial temperature is required. Secondly, the dipole trap experiments are simulated. Therein, the ions are transferred from the effective potential of the RF trap into the dipole trap, in which they are exposed to recoil heating. Algorithms from Ref. 27 were used to integrate the equation of motion (Runge-Kutta-Fehlberg (4,5) method) and to generate pseudo-random numbers from different distributions.

For the Doppler cooling simulation the potential consists of the superposition of the time-averaged RF potential and the static electric potential. For both individual potentials one principal axis coincides with the $z$-axis, while the other principal axes $\hat{e}_{x'}$ and $\hat{e}_{y'}$ can be rotated by an angle $\alpha$ referred to the $x$- and $y$-axis, respectively. (In general, the rotation angle of the time-averaged RF potential will be different from that of the static electric potential. The rotation angle of the total potential will by somewhere inbetween. This is neglected in the following.) The relation $\omega_{x'}^2 + \omega_{y'}^2 + \omega_z^2 = 0$ must hold for the frequencies of the static electric potential. For the time-averaged RF potential the relation $\omega_{>}^{(RF)} - \omega_{<,1}^{(RF)} - \omega_{<,2}^{(RF)} = 0$ holds, where $\omega_{>}^{(RF)}$ denotes the largest of the three frequencies $\omega_{x/y/z}^{(RF)}$ and $\omega_{<,1/2}^{(RF)}$ the two smaller ones. A small axial confinement $\omega_z^{(RF)} \sim 2\pi \times$ kHz is due to the geometry of the electrodes. The axial frequency of the total potential is still well described by $\omega_z$ (the correction due to $\omega_z^{(RF)}$ can be neglected) and the radial frequencies are split by few $2\pi \times$ kHz. Unfortunately, both the exact splitting of the radial frequencies as well as the angle $\alpha$ are experimentally difficult to access in the current setup. The reliability of numerical calculations of the potentials of the RF trap is restricted due to the difficulties discussed in Sec. I A.

In the Doppler cooling simulation both superimposed Doppler cooling lasers DL (with detuning of $\Delta_{DL} \approx -\Gamma/2$ from the $S_{1/2} \leftrightarrow P_{3/2}$ transition to cool to a low temperature) and DLdet ($\Delta_{DLdet} \approx -11\Gamma$ to re-cool occasionally heated ions) are included. They are focussed onto the ion under an angle of $22.5°$ to the $z$-axis ($\hat{k}_{DL} = (0, \sin(22.5°), \cos(22.5°))$, see Fig. 1) and have waist radii around $90\,\mu$m. The polarization of the laser beams is the same and we assume a $\pi$-polarization in the simulation, as it is the main component.

Each time step of the simulation consists of

1. a propagation of the ion for a duration $\Delta t$ in the potential,
2. an absorption or induced emission of a photon from/into each Doppler cooling beam with a certain probability, and
3. a spontaneous decay with isotropic emission of a photon (for excited states $i$) with a certain probability.

The results of different Doppler cooling simulations are shown in Fig. 13. The powers of the Doppler cooling lasers $P_{DL/DLdet}$ corresponding to Fig. 10 are larger (on-resonance saturation parameter for the laser DL $s_0 \approx 0.3$) than the powers corresponding to Fig. 11 and Doppler cooling becomes slightly less efficient. The light pressure excerts a mean force of $\sim 3 \times 10^{-20}$ N onto the ion and leads to a significant displacement $\Delta z = 8.8\,\mu$m. The resulting probability distributions deviate only lightly from the Boltzmann distributions of the three-dimensional harmonic oscillator, which are shown for comparison.

In Fig. 14 the results of Doppler cooling simulations are compared for different splittings $\omega_{x'} - \omega_{y'}$ and angles $\alpha$, respectively. Doppler cooling becomes more efficient for large angles $\alpha$ and large splittings of the radial frequencies. While the impact of the angle is large, different splittings lead to only a small change as long as they are larger than $\sim 2\pi \times 1$ kHz. Hence, the result of the Doppler cooling is comparatively insensitive to any (plausible) estimates of the splitting of the radial frequencies.

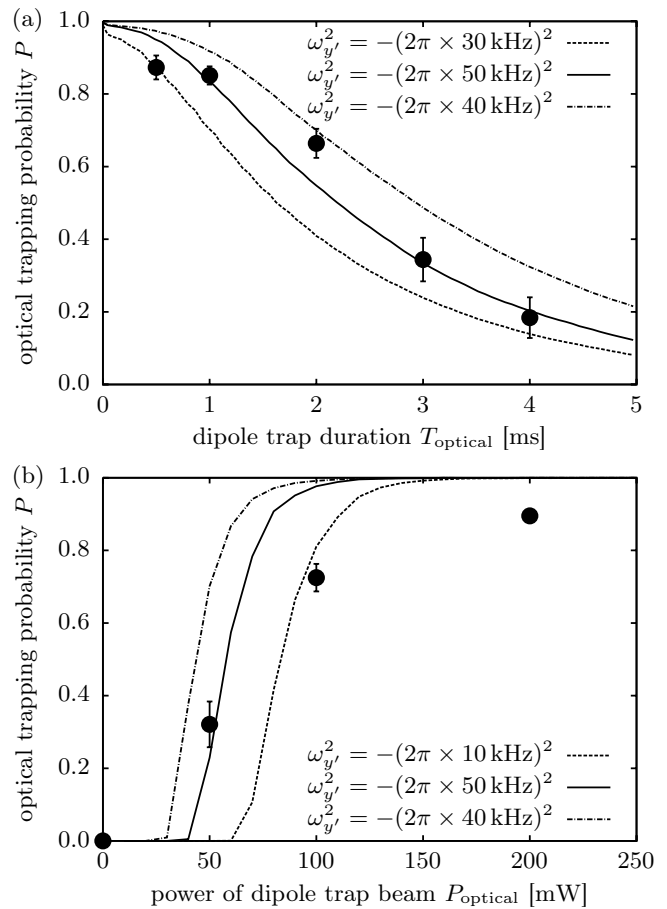

Figure 15. Monte-Carlo simulations of the dipole trap experiments (cmp. Figs. 10 and 11) for an angle $\alpha = 35°$ and different frequencies $\omega_{y'}^2$. The initial values for the position, velocity and internal state are obtained from the respective results of the Doppler cooling simulations depicted in Fig. 13. The points represent the measured optical trapping probabilities and the error bars the standard deviation of the mean. The Monte-Carlo simulations agrees with the measured trapping probabilities for the depicted parameters.

These dependencies of the Doppler cooling results are consistent with theory: Reducing the power of a Doppler cooling beam, reduces the mean energy of the particle, as long as external heating effects remain negligible. As the cooling beams are sent in along one direction only, they must not be (close to) perpendicular to any of the principal axes of the trap potential for Doppler cooling to be effective in all dimensions. This also implies that the frequencies of the total potential must be pairwise different.

Next, the dipole trap experiments are simulated. The initial conditions (position, velocity, and internal state) are picked from results of the Doppler cooling simulation. Initially, the potential consists of the one used during Doppler cooling superimposed with the dipole trap potential $U_{\text{optical}}(\vec{r})$ as in Sec. II B, but considering the slight asymmetry in the waist radii in horizontal and vertical direction ($w_{0,h} = 0.7\,w_{0,v} = 0.7\,w_0$). In the experiment the dipole trap beam is focussed on the ion by observing a decrease of the Doppler cooling fluorescence due to the AC Stark shift caused by the dipole trap beam. The power of the dipole trap beam and thus the decrease of the fluorescence is kept as small as possible, such that the dipole trap beam is effectively focussed onto the ion displaced by the light pressure caused by the Doppler cooling beams. That is why the dipole trap potential in the simulation is also displaced by $\Delta z$ from the center of the potentials of the RF trap. The frequencies $\omega^{(\text{RF})}_{x'/y'/z}$ of the time-averaged RF potential are linearly ramped down to zero within the first 50 µs of the simulation. Both $\omega_{y'}$ (or $\omega_{x'}$) and $\alpha$ are free parameters as in the Doppler cooling simulations.

The trapped ion randomly absorbs photons of the dipole trap laser with the scattering rate $\Gamma_s(\vec{r})$. As $\Gamma_s(\vec{r}) \ll \Gamma$, every absorption is followed by an immediate, isotropic spontaneous decay. The ion will be considered as lost from the trap, if its distance to the origin exceeds several waist radii $w_0$. Hence, we can be sure that it left the trapping region of the dipole trap and within roughly 10 µs $\ll T_{\text{optical}}$ it will also be accelerated out of the region, where the RF trap could re-capture it. The trapping probability $P$ for different parameters $T_{\text{optical}}$ and $P_{\text{optical}}$ is determined as the fraction of ions that were not lost.

In Fig. 15, the results are illustrated for an angle $\alpha = 35°$ and different frequencies $\omega_{y'}$. The best agreement with the measured optical trapping probabilities is obtained for $\omega_{y'}^2 \approx -(2\pi \times 50\,\text{kHz})^2$ (and similarly for $\omega_{y'}^2 \approx -(2\pi \times 35\,\text{kHz})^2$, which is not shown). For both simulations the maximal trapping probability is reached for $\omega_{y'}^2 \approx -(2\pi \times 40\,\text{kHz})^2$. This is consistent with the expectations from Sec. II B. (Note, however, that the potential has been simplified in these analytic calculations and they strictly hold for $\alpha = 0°$ only.) The simulation of the lifetime measurement (Fig. 15a) shows that the comparatively large displacement $\Delta z > w_0$ of the ion does not prevent it from being trapped successfully. The comparatively slow ramp-down of the RF potential allows the ion to be displaced mainly along the propagation direction of the dipole trap beam and loose most of the potential energy related to the displacement $\Delta z$ from the center of the static electric potential. The simulation of the optical trapping probability $P$ versus power of the dipole trap beam $P_{\text{optical}}$ (Fig. 15b) shows a comparatively sharp increase for small powers and $P$ almost reaches 1. The deviation from the measured probabilities can most likely be explained by beam pointing instabilities in the experiment, that smoothen the sharp increase and reduce the maximal trapping probability.

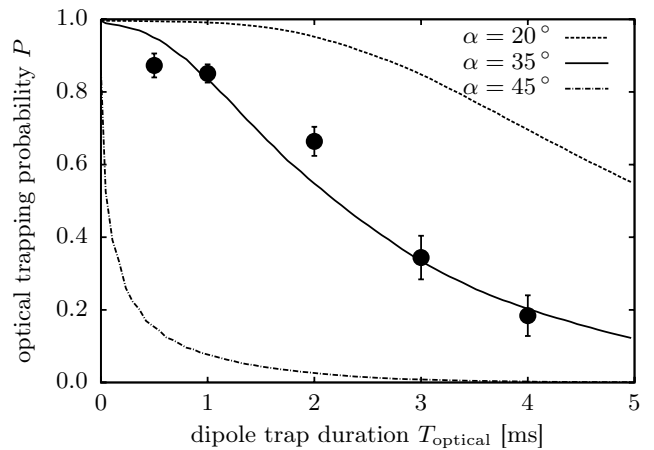

Figure 16. Monte-Carlo simulations of one dipole trap experiment (cmp. Fig. 10) for a frequency $\omega_{y'}^2 = -(2\pi \times 50\,\text{kHz})^2$ and different angles $\alpha$ between principal axes of the static electric potential and the axes of the coordinate system. The initial values for the position, velocity and internal state are obtained from the same Doppler cooling simulation of Fig. 13 for all three curves. Note that this allows for a better comparison of the influence of $\alpha$ on the dipole trap experiment, but assumes a better Doppler cooling result for small angles than predicted by the simulations in Fig. 14. The trapping probability shows a strong dependence on the angle $\alpha$.

These beam pointing instabilies are mainly caused by the convection near the warm magnetic field coils of the quantization field.

Within a certain range of parameters both the Doppler cooling and the dipole trap experiments are quite insensitive to the actual values of the free parameters. The lifetime experiment, for example, is consistent within the error bounds of the experimental parameters with $\omega_{y'}^2 = -(2\pi \times [30\,\text{kHz}\ldots 60\,\text{kHz}])^2$ at an angle $\alpha = 35°$. However, for frequencies $\omega_{y'}^2$ outside this range the trapping probability is significantly reduced or even zero. The optical trapping probability $P$ drops for some specific angles $\alpha$ towards zero (see Fig. 16 for $\alpha = 45°$) or is enhanced (see Fig. 16 for $\alpha = 20°$), as long as the same Doppler cooling efficieny is assumed for all simulations.

The Monte-Carlo simulation still does not account for some effects that most probably also cause a decrease of the optical trapping probability, e.g. micromotion during the Doppler cooling and ramp-down of the RF potential, pointing instabilities of the dipole trap beam including an overall drift, residual stray electric fields, an arbitrarily rotated static electric potential (including the $z$-principal axis), or a reduction of the peak intensity of the dipole trap beam deviating from a perfect mode due to diffraction, abberations, etc. Still, it overcomes the most severe limitations of the analytic models. Its results are in agreement with the measured trapping probabilities and support recoil heating as the relevant heating source in the dipole trap experiments. Additionally, the Monte-Carlo simulation allows to reveal the large impact of the radial frequencies $(\omega_{x'}^2, \omega_{y'}^2)$ of the static electric potential and of the angle $\alpha$ of the principal axes. The investigations also show that it is unlikely that the trapping probability can be significantly improved in the current setup, at least, without cooling of the ion in the dipole trap. The Monte-Carlo simulation is expected to also be a useful tool for future experiments to allow for both quantitative predictions and analyses of measurements.

## III. OUTLOOK

Our experiment leaves room for various technical improvements. First of all, the recoil heating rate must be reduced to allow for longer trapping durations. This can be accomplished by a dipole trap beam with larger detuning $\Delta$ and higher laser power. Laser systems as described in Ref. 28, a laser at 1120 nm [1, 29], or systems at other prominent visible or infrared wavelengths (e.g. 532 nm or 1064 nm) are established candidates for this purpose and available as turn-key systems. These wavelengths provide further advantages compared to the ultraviolet wavelength used in our experiment: Optical fibres are available that allow for short beam paths in air and hence stable beam pointing and beam shapes. In contrast, the length of the beam path in the presented experiments amounts to several metres and the beam shape of the dipole trap suffers from the strong astigmatism of the second SHG. Stray electric fields drifting on short timescales due to charges produced by stray laser light via the photoelectric effect should be mitigated for long wavelengths. Obviously, a reduction of the distance of the windows of the vacuum chamber to the trap would ease to achieve small and robust waist radii.

The RF trap has proven to be a useful tool to allow for the sensing and compensation of stray electric fields and loading an ion into the dipole trap, although in principle an ion could also be created by ionizing an already trapped neutral atom. The RF trap used in the presented experiments was originally optimized for quantum simulation experiments [10, 11] requiring high secular frequencies. This results in small distances of the ion to the electrodes. As opposed to that, the dipole trap experiments would benefit from a trap optimized for small secular frequencies with large electrode–ion distances. The latter would in addition ease the alignment of high-power dipole trap beams with low amount of stray light. If a linear RF trap is used, an improved setup shall allow to align the (or one of the) dipole trap beam(s) with the axis of the RF trap to become more insensitive to radially defocussing DC electric fields. Alternatively, as long as only one or few ions have to be transferred into an optical trap, it might be advantageous to use an electrode geometry that provides an RF confinement in all three dimensions (e.g., a larger version of the stylus trap in Ref. 30). In such a trap DC fields would only be required to compensate stray fields in contrast to a linear RF trap, where these fields also have to provide the axial confinement. As a result, the DC fields and thus their defocussing effect could be kept weak. An additional transfer of the ion into an experimental zone could mitigate the problems caused by the coating of the electrodes with the neutral atoms from the atomic oven (cmp. Sec. I A). Ablation loading [31] might be an alternative.

Displacements of the ion(s) due to light pressure can be avoided by using counter-propagating Doppler cooling beams, which becomes more important for further reduced secular frequencies. Different ion species allow for lower Doppler cooling temperatures, however, at the expense of a more complicated level scheme and thus a more involved optical trapping scheme. A better pre-cooling before transferring the ion into the dipole trap, e.g., by cooling it to the motional ground state [14], might allow for shallower optical potentials. In order to enhance the lifetime of an ion in the dipole trap, cooling techniques used for atoms in optical traps should also be applicable to ions. For example, sideband or Sisyphus cooling could be implemented and cavity cooling, as demonstrated in Ref. 32, might not even have to affect the electronic coherences. Alternatively, ions could be sympathetically cooled by atoms within a magneto-optical trap or a Bose–Einstein condensate [33].

We refer the reader to Ref. 1 for an extended discussions of possible applications. In the near future, cold-collision experiments with atoms and ions could benefit from the effectively micromotion-free trapping of the ion in an optical trap [34]. Currently, such experiments are implemented as hybrid setups with an ion confined in an RF trap and atoms in a magneto-optical trap [35] or a Bose–Einstein condensate [33, 36]. As the ion is subject to RF-driven micromotion, the system of the ion and the atoms experiences RF-induced heating and the collision energy in these experiments is limited to the order of millikelvins. Also, superimposing an optical lattice with ions trapped in an RF trap for quantum simulation experiments is in reach [37, 38]. On longer timescales, multiple ions in sparsely populated optical lattices or one (few) ion(s) within an optical lattice filled with neutral atoms might be exploited for quantum simulations.

## ACKNOWLEDGEMENT

This work was supported by the Max-Planck-Institut für Quantenoptik (MPQ), Max-Planck-Gesellschaft (MPG), Deutsche Forschungsgemeinschaft (DFG) (SCHA 973/1-6), the European Commission (The Physics of Ion Coulomb Crystals: FP7 2007–2013. grant no. 249958) and the DFG Cluster of Excellence "Munich Centre for Advanced Photonics". We thank Hector Schmitz, Robert Matjeschk, and Michael Szelong for preliminary work and Dietrich Leibfried, Karim Murr, and Roman Schmied for helpful discussions.